# Accurate, Fast and Lightweight Clustering of *de novo* Transcriptomes using Fragment Equivalence Classes

Avi Srivastava [1,*], Hirak Sarkar [1,*], Laraib Malik [1], Rob Patro[1]

[1] Department of Computer Science, Stony Brook University Stony Brook, NY 11794-2424

## Abstract

**Motivation**: *de novo* transcriptome assembly of non-model organisms is the first major step for many RNA-seq analysis tasks. Current methods for *de novo* assembly often report a large number of contiguous sequences (contigs), which may be fractured and incomplete sequences instead of full-length transcripts. Dealing with a large number of such contigs can slow and complicate downstream analysis.
**Results**: We present a method for clustering contigs from *de novo* transcriptome assemblies based upon the relationships exposed by multi-mapping sequencing fragments. Specifically, we cast the problem of clustering contigs as one of clustering a sparse graph that is induced by equivalence classes of fragments that map to subsets of the transcriptome. Leveraging recent developments in efficient read mapping and transcript quantification, we have developed `RapClust`, a tool implementing this approach that is capable of accurately clustering most large *de novo* transcriptomes in a matter of minutes, while simultaneously providing accurate estimates of expression for the resulting clusters. We compare `RapClust` against a number of tools commonly used for *de novo* transcriptome clustering. Using *de novo* assemblies of organisms for which reference genomes are available, we assess the accuracy of these different methods in terms of the quality of the resulting clusterings, and the concordance of differential expression tests with those based on ground truth clusters. We find that `RapClust` produces clusters of comparable or better quality than existing state-of-the-art approaches, and does so substantially faster. `RapClust` also confers a large benefit in terms of space usage, as it produces only succinct intermediate files — usually on the order of a few megabytes — even when processing hundreds of millions of reads.
**Availability**: RapClust, is implemented in Python, and is available as open source software at https://github.com/COMBINE-lab/RapClust under a BSD with attribution license.
**Contact**: rob.patro@cs.stonybrook.edu

## 1 Introduction

RNA-seq has proven to be a crucial tool in the study of transcriptomes, allowing scientists to probe differential expression under various conditions, like response to changing stimuli (27) and disease states (25). It also allows for the discovery of novel alternative splicing events within annotated genes as well as the discovery of previously unannotated genes. One of the particularly useful properties of high-throughput RNA-seq data is that it can be used to directly assemble the transcriptome of organisms, even when a reference genome is not available. *de novo* transcriptome assembly solves this problem by assembling RNA-seq reads directly into contigs. Popular assemblers, like Trinity (7), Oasis (21) and Trans-ABySS (20), employ a wide range of algorithms and heuristics to provide contigs that approximate the full-length transcripts present in the underlying sample. Given the cost and difficulty that may be required for genomic assembly, *de novo* transcriptome assembly offers a compelling alternative when one is interested primarily in studying an organism's transcriptome.

Despite the advanced techniques employed by many modern *de novo* transcriptome assemblers, the resulting assemblies often contain a large number of contigs which do not represent full-length transcripts. These incomplete contigs may result from fractured assemblies, incomplete assemblies due to lack of coverage, errant assembly of chimeric transcripts, or a host of other errors (22). Though improved computational methods can reduce the prevalence of such errors, the data itself is often insufficient to guarantee deterministic recovery of all expressed transcripts. The fractured and incomplete nature of such *de novo* assemblies can confound downstream analysis. For example, Davidson and Oshlack (5) argue that the large number of contigs that often result from *de novo* transcriptome assembly can greatly reduce the statistical power of differential expression analysis. This results, in part, from the need to correct for testing the many additional hypotheses that arise from considering the large number of assembled contigs (which is likely much greater than the actual number of transcripts expressed in the sample). Moreover, the sequence-similar contigs generated by *de novo* assemblers typically result in a high fraction of ambiguous, multi-mapping reads. This ambiguity is challenging to resolve, but must be accounted for when performing differential expression analysis at the contig level.

---
[*]These authors contributed equally to this work.

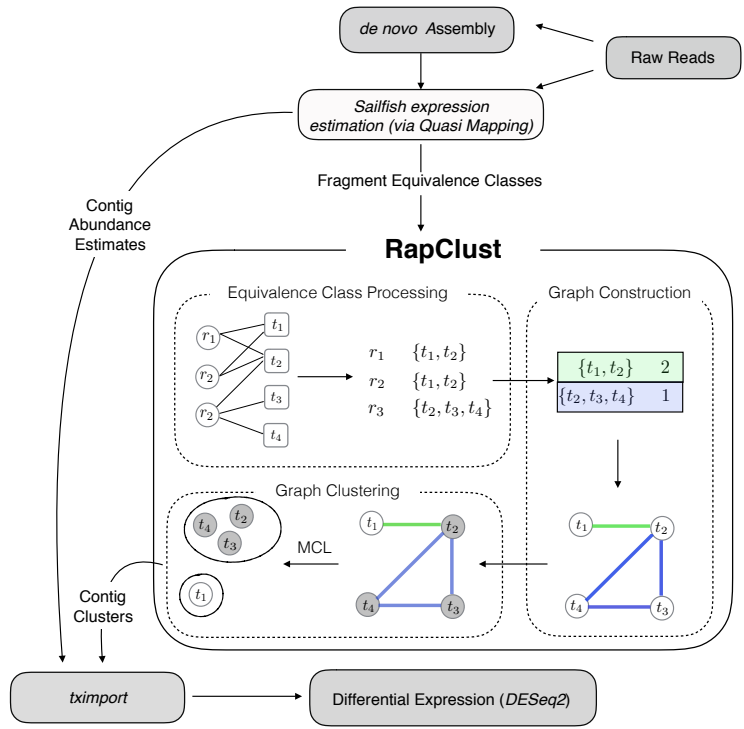

Figure 1: An overview of the `RapClust` pipeline. Fragment equivalence classes are computed using *quasi-mapping*, and these classes are used for both contig-level expression quantification and generation of the mapping ambiguity graph. The mapping ambiguity graph is partitioned using `MCL`. The resulting clusters can then be used for downstream analysis (e.g. differential expression).

Common pipelines for studying differential gene expression across experimental conditions first align the RNA-seq reads back to the assembled contigs. Then, they use transcript-level expression estimation tools, such as RSEM (10), to account for the high degree of multi-mapping ambiguity that results from the substantial sequence similarity between related contigs. To further improve expression estimates, contigs that have high similarity (e.g. that are very sequence-similar or that have many overlapping reads aligning between them) are clustered together as putative transcripts or contigs of the same gene. Statistical methods, such as those discussed in (9, 23), are then used to identify contigs (or clusters of contigs) that are likely to be differentially expressed across conditions. Clustering of contigs into putative genes can be a crucial step in this analysis, since performing differential expression analysis at the level of clusters reduces the multiple hypothesis testing burden, which can be high in *de novo* transcriptomes, owing to the potentially large number of assembled contigs. Additionally, aggregating contigs into such groups can improve the robustness of expression estimation and hence the accuracy of differential expression analysis. We note that, if one requires transcript-level differential expression, such a clustering procedure may not always be useful. However, for many analyses, it is beneficial.

Although clustering may help improve the accuracy of differential expression results, it should be designed to account for the multiple sources of sequence similarity that appear in the assembly. For example, paralogs should ideally be placed in separate clusters, while isoforms of the same gene should be co-clustered. The effects of different clustering approaches in the context of analyzing *de novo* transcriptomes has previously been explored in depth, e.g. by Davidson and Oshlack (5), which largely inspired the current work. Unfortunately, the approaches tend either to have high computational requirements (mainly due to their need to align 10s or 100s of millions of reads back to the assembly), or can yield clusters that may poorly reflect the true relationship between contigs and genes (11). In this paper, we present `RapClust`, a tool for clustering contigs in *de novo* transcriptome assemblies. `RapClust` achieves comparable accuracy to state-of-the-art transcriptome clustering methods, while being much faster. `RapClust` works in conjunction with the `Sailfish` tool, which is already capable of quickly and accurately producing contig-level abundance estimates. It uses the fragment equivalence classes computed by `Sailfish` to derive accurate and biologically meaningful clusters at only a marginal extra cost, beyond what is required for quantification.

An overview of the `RapClust` pipeline is given in Figure 1. `RapClust` requires only a transcriptome assembly, the raw sequencing reads, and a description of the experimental design; it yields an accurate contig clustering, along with both contig and cluster-level expression estimates. `RapClust` is agnostic to the choice of the underlying *de novo* assembler, and can be used with popular tools such as Trinity (7) or Oases (21). *Quasi-mapping*, a recently introduced (26) and fast alternative to read alignment, is used to map the reads to the assembled contigs. The multi-mapping structure of the sequencing reads with respect to the transcriptome is encoded in the form of *fragment equivalence classes*, as discussed in Section 2.1. These equivalence classes induce a mapping ambiguity graph as detailed in Section 2.2 (the notion of the fragment ambiguity graph has previously proven useful e.g. in the context of re-estimating transcript abundances after updates to an annotation (19)). After some post-processing, the resulting graph is clustered using `MCL` (31). The computed clusters represent a putative contig-to-gene mapping, which can then be used to aggregate the contig-level expression estimates derived by `Sailfish`. These cluster-level expression estimates can then be used for downstream analyses like differential expression testing.

## 2 Methods

In this work, we make explicit the connection between the successful approach of *de novo* transcriptome clustering presented in (5), and the concept of equivalence classes over fragments that has enabled, in part, a new class of ultra-fast methods for transcript-level quantification from RNA-seq data (3, 17, 18). The notion of fragment equivalence classes, as a means of factorizing the likelihood function used in transcript-level quantification, was originally introduced by Nicolae et al. (14) and Turro et al. (30) (though a related factorization was used somewhat earlier by Jiang and Wong (8)). Substantial speed improvements were obtained when traditional alignment of fragments was replaced with much faster procedures like k-mer counting (17), lightweight-alignment (18), pseudoalignment (3) and *quasi-mapping* (26). All of the ultra-fast transcript quantification tools mentioned above use these fast alternatives to alignment together with the notion of fragment equivalence classes. Recently, Ntranos et al. (16) showed that fragment equivalence class counts (called transcript-compatibility counts therein) can be effectively used to cluster cells in single-cell RNA-seq experiments. In our current work, by drawing on these connections, and by focusing on the rich information exposed by fragment equivalence classes, we frame the transcript clustering problem in the context of the mapping ambiguity graph induced by these equivalence classes. This allows for the rapid clustering of contigs, on the basis of both sequence and expression similarity, using only small intermediate space.

## 2.1 Computing Equivalence Classes via Quasi-mapping

The concept of *quasi-mapping*, which provides information about the transcripts, positions and orientations from which a fragment has possibly originated, but not the base-to-base alignment by which the fragment corresponds to the transcript, has recently been introduced in Srivastava et al. (26). There, it is suggested that *quasi-mapping* may be adopted as a much-faster alternative to fragment alignment when the base-to-base alignments are not required for the task being performed. Srivastava et al. describe an efficient implementation *quasi-mapping* in the tool `RapMap`, and demonstrate how integrating *quasi-mapping* in the `Sailfish` software for transcript-level quantification led to considerable improvements in accuracy and speed. Here, we rely on *quasi-mapping* to allow for very fast and accurate determination of fragment equivalence classes, which is crucial to the approach we adopt below. We note that, though *quasi-mapping* does not compute a nucleotide-level alignment, it is sensitive to even small differences in related reference sequences. Thus, it can accurately map fragments to e.g. the appropriate paralog, even if the fragment contains only a single SNP differentiating the alternative reference sequences. We refer the reader to (26) for details of the *quasi-mapping* concept and the particular algorithm used by `RapMap`.

In this paper, we rely heavily on the concept of fragment equivalence classes. We define an equivalence relation over fragments, based on the set of transcripts to which they map. The set of fragments related under this definition constitutes a fragment equivalence class. Let $\mathcal{M}(f_i)$ be the set of transcripts to which fragment $f_i$ maps, and let $\mathcal{M}(f_j)$ be the set of transcripts to which fragment $f_j$ maps. We say that $f_i \sim f_j$ if and only if $\mathcal{M}(f_i) = \mathcal{M}(f_j)$. Consequently, a fragment equivalence class is a set of fragments such that, for every pair $f_i$ and $f_j$ in the class, $f_i \sim f_j$. An equivalence class can be uniquely labeled based on the set of transcripts to which the fragments contained in this class map. We define the label of equivalence class $[f_i] = \{f_j \in \mathcal{F} \mid f_j \sim f_i\}$ as $\text{lab}([f_i])$. It is important to remember that, though the label consists of transcript names, the equivalence relation itself is defined over sequenced fragments and not transcripts. Finally, in addition to a label, we denote the count of each equivalence class $C_i$ by $\text{count}(C_i)$; this is simply the number of equivalent fragments in $C_i$.

## 2.2 Quantification and Graph Determination

The fragment equivalence classes, as described above, are already computed internally by `Sailfish` (17). We have modified `Sailfish` to write these equivalence classes to a file once quantification is complete (this behavior is enabled with the `-dumpEq` flag). This yields, for each *sample*, a collection of equivalence classes, along with their associated labels and counts. To construct the complete mapping ambiguity graph of the *experiment* we need to aggregate these equivalence classes over all of the processed samples. In fact, this aggregation is relatively simple since the labels of fragment equivalence classes are deterministic and stable across samples (i.e. they depend only on the underlying transcriptome). We generate a single collection $\mathscr{C}$ of equivalence classes by merging the classes $\mathscr{C}_1, \ldots, \mathscr{C}_M$, from all samples, where $M$ is the number of samples. Here, $\mathscr{C}$ contains the union of equivalence classes from $\mathscr{C}_1, \ldots, \mathscr{C}_M$, and classes that appear in more than one sample of $\mathscr{C}_1, \ldots, \mathscr{C}_M$ simply have their respective read count summed. The time and space requirement for this aggregation algorithm is linear in the size of input.

For a given experiment, the collection $\mathscr{C} = \{C_1, C_2, \ldots, C_k\}$ of equivalence classes induces a weighted, undirected, mapping ambiguity graph $G = (V, E)$. Here, $V = T$, where $T$ is the set of transcripts in the original transcriptome — and $E = \{\{t_i, t_j\} \mid \exists C_\ell \in \mathscr{C} \text{ where } \{t_i, t_j\} \subseteq \text{lab}(C_\ell)\}$ — that is $t_i$ and $t_j$ are connected by an edge if and only if they both appear in the label of at least one equivalence class. For a given edge $\{t_i, t_j\}$, its weight is given by $w(t_i, t_j) = N_{ij}/\min(N_i, N_j)$, where

$$N_{ij} = \sum_{\substack{C_\ell \in \mathscr{C} \mid \\ \{t_i,t_j\} \subseteq \text{lab}(C_\ell)}} \text{count}(C_\ell), \quad N_i = \sum_{\substack{C_\ell \in \mathscr{C} \mid \\ t_i \in \text{lab}(C_\ell)}} \text{count}(C_\ell) \text{ and } N_j = \sum_{\substack{C_\ell \in \mathscr{C} \mid \\ t_j \in \text{lab}(C_\ell)}} \text{count}(C_\ell).$$

While the samples we process from a *de novo* RNA-seq experiment may contain 10s to 100s of millions of fragments, the number of nodes in $G$ is determined by the number of contigs. Further, the number of edges is bounded by the *complexity* of the transcriptome (i.e. the degree of alternative splicing and paralogy in the underlying transcriptome), and, therefore, mostly independent of the number of fragments processed.

For the transcriptomes and samples we analyze in this paper, the number of equivalence classes never rises above a few hundred thousand, and is typically orders of magnitude smaller than the number of fragments (see table 1 for the number of equivalence classes in the different data sets). Thus, if the equivalence classes can be computed efficiently from the transcriptome and sequencing data, then the mapping ambiguity graph can be constructed efficiently in terms of time and space.

### 2.3 Processing and Partitioning the Mapping Ambiguity Graph

Once the mapping ambiguity graph *G* is constructed, it is filtered, as described below, to yield a graph $G'$. $G'$ is then clustered using an off-the-shelf graph clustering algorithm. Currently, `RapClust` employs two simple filters. The first filter removes nodes from *G* that have fewer than some nominal threshold of read support over all samples in an experiment. We adopt the cutoff used by (5), and remove any contig with 10 or fewer mapped reads from the mapping ambiguity graph.

The second filter, also adopted from (5), is used to remove edges between pairs of contigs that are likely to arise from paralogous genes. Specifically, this filter tests the hypothesis that the constant of proportionality between the number of reads mapping to $t_i$ and $t_j$ does not vary (by a statistically significant amount) across conditions. This is done by testing the hypothesis ($H_0$) that the constant of proportionality remains constant across conditions versus the hypothesis ($H_1$) that it does not. The log-likelihoods of the competing hypotheses $H_0$ and $H_1$ are computed according to eqs. (1) and (2) respectively. A likelihood ratio test is performed, and edges $\{t_i, t_j\}$ from *G* where $2(\ell_1 - \ell_0) > 20$ are removed (we refer the reader to (5) for a justification of the cutoff used in the likelihood ratio test). This test, of course, makes some simplifying assumptions, since isoforms of the same gene might exhibit behavior consistent with $H_1$ (e.g. if isoform switching occurs between conditions). However, we found that this filter correctly separates transcripts from paralogous genes more often than it incorrectly separates transcripts of the same gene. Thus, applying this filter leads to a slight increase in `RapClust`'s precision and a typically smaller decrease in its recall.

$$\ell_0 = \sum_c \left[ \left(X_i^c \cdot \log\left(r_{ij}\mu_j^c\right)\right) - \left(r_{ij}\mu_j^c\right) \right] + \left[ \left(X_i^c \cdot \log\left(\mu_j^c\right)\right) - \left(\mu_j^c\right) \right] \tag{1}$$

$$\ell_1 = \sum_c \left[ \left(X_i^c \cdot \log\left(r_{ij}^c X_j^c\right)\right) - \left(r_{ij}^c X_j^c\right) \right] + \left[ \left(X_i^c \cdot \log\left(X_j^c\right)\right) - \left(X_j^c\right) \right] \tag{2}$$

where

$$r_{ij}^c = \frac{X_i^c}{X_j^c}, \ r_{ij} = \frac{\sum_c X_i^c}{\sum_c X_j^c}, \text{ and } \mu_j^c = \frac{X_i^c + X_j^c}{1 + r_{ij}},$$

and $X_i^c$ denotes the number of reads mapping to contig *i* under the *j*[th] condition (summed over all replicates of a condition for simplicity).

After applying both of these filters in sequence, we obtain the final processed graph $G'$, which is then clustered using `MCL` (31). For the sake of simplicity, and to avoid an unnecessary dependence on parameters, we generated all the clusterings in this paper using `MCL`'s default parameters, and applying no additional cutoff or modification to the edge weights.

## 3 Results

To analyze the performance of `RapClust`, we have benchmarked its running time, space usage, and accuracy against `Corset` (5) and `CD-HIT` (6, 11) (here, we consider `CD-HIT -EST`). Note that `CD-HIT` does not provide any quantification results. Therefore, in Section 3.3, we considered the clustering computed by `CD-HIT`, but estimated the expression of those clusters by aggregating the contig-level expression estimates computed by `Sailfish`. Testing was performed on 3 datasets, human primary lung fibroblast samples, with and without a small interfering RNA (siRNA) knock down of HOXA1 (29) (Gene Expression Omnibus accession GSE37704), yeast grown under batch and chemostat conditions (15) (Sequence Read Archive (SRA) accessions SRR453566 to SRR453571) and male and female chicken embryonic tissue (1) (SRA accession SRA055442). For each of these data sets, we performed clustering of Trinity (7) *de novo* assemblies, which were obtained from (4).

All experiments were performed on a 64-bit Linux server, running Ubuntu 14.04, with 4 hexacore Intel Xeon E5-4607 v2 CPUs (with hyper-threading) running at 2.60GHz and 256GB of RAM. Wall-clock time was recorded using the Unix `time` command. Table 1 gives a brief description of the input data.

Table 1: Summary statistics for the transcriptomes and experimental samples on which the experiments were carried out, as well as the average number of fragment equivalence classes and the size of the resulting mapping ambiguity graph.

|  | Yeast | Human | Chicken |
| --- | ---: | ---: | ---: |
| # contigs | 7353 | 107,389 | 335,377 |
| # samples | 6 | 6 | 8 |
| Total (paired-end) reads | ~36,000,000 | ~116,000,000 | ~181,402,780 |
| Avg # eq. classes (across samples) | 5197 | 100,535 | 222,216 |
| # edges in mapping ambiguity graph | 6195 | 212,481 | 2,063,524 |

## 3.1 Time and Space Requirements

Here, we report, for each clustering method, the time required to perform the clustering as well as the total size of the intermediate and result files written to disk. It is important to note that, unlike `RapClust` and `Corset`, `CD-HIT` neither counts reads nor performs quantification. In order to use the clusters resulting from `CD-HIT` in a typical analysis (e.g. quantification and differential expression testing), one would need to either align reads to the `CD-HIT` clusters, or perform quantification on these clusters using the sequenced reads, both of which would add to the time and disk space required.

For each transcriptome, the time reported for `Corset` is the sum of the time required to trim the reads using `Trimmomatic` (2), align the reads using `Bowtie`, and cluster the contigs using the resulting BAM files (this adopts the protocol and parameters suggested in the `Corset` documentation). For `RapClust`, the time reported is the time required to run `Sailfish` (v0.9.1) on all the samples, plus the time required to generate, filter and cluster the mapping ambiguity graph. `Sailfish` is run with the `-dumpEq` option to write to disk the equivalence classes computed during the quantification of each sample. `Trimmomatic`, `Bowtie`, and `Sailfish` were each run with 4 threads. For `CD-HIT`, only the time required to run `CD-HIT` is reported. To determine the disk space required for analysis of each transcriptome using `Corset`, we sum the size of the BAM files produced by `Bowtie` and the "count" and "cluster" files produced by `Corset`. For `RapClust`, we determined the required intermediate disk space by summing the sizes of the `Sailfish` quantification directories and the "graph" and "cluster" files. Since both methods require the same input in terms of the assembled transcriptome and set of reads, we don't count these toward the space requirements. The complete timing results (from assembly and raw reads to computed clusters) for `RapClust` and `Corset` are presented in Table 2. The times required for *just* the clustering steps of `RapClust` and `Corset`, as well as the time required by `CD-HIT`, are reported in Table 3. We choose to report these results separately to highlight the fact that, since `CD-HIT` only performs clustering, if one wishes to carry out expression analysis on the clusters computed by `CD-HIT`, she would additionally have to either align the sequenced fragments or perform contig-level abundance estimation on the samples, which could take much longer.

Table 2: `RapClust` is *substantially* faster than `Corset`, and requires only a small fraction of the intermediate disk space used by `Corset`. The majority of space savings for `RapClust` come as a result of avoiding alignment and generation of the intermediate BAM files. In addition to that, the percentage of reads mapped using `RapClust` is much higher. Originally, the percentage of mapped reads for `RapClust` was even higher than what is reported here, but we subsequently modified our default behavior to discard orphan mappings to be consistent with the behavior of the alignments provided to `Corset`.

|  | Yeast | | Human | | Chicken | |
| --- | ---: | ---: | ---: | ---: | ---: | ---: |
|  | RapClust | Corset | RapClust | Corset | RapClust | Corset |
| Time(min) | 5.12 | 37.25 | 22.67 | 211.67 | 64.18 | 453 |
| Space(Gb) | 0.005 | 5.7 | 0.092 | 22 | 0.49 | 145 |
| % of reads | 88.17 | 62.32 | 93.04 | 77.94 | 88.80 | 60.99 |

Table 3: The overall time taken for `Corset` and `RapClust` are dominated by the time taken for alignment and quantification respectively. However, when we consider just the time required for clustering (i.e. after alignments have been generated for `Corset` and after the fragment equivalence classes have been generated for `RapClust`), we observe that `RapClust` and `CD-HIT` are considerably faster than `Corset`. As `Corset`'s clustering phase is single-threaded, we provide times for all methods in this table using only a single thread. (RC = `RapClust`, CD = `CD-HIT`, CT = `Corset`)

|  | Yeast | | | Human | | | Chicken | | |
| --- | --- | --- | --- | --- | --- | --- | --- | --- | --- |
|  | RC | CD | CT | RC | CD | CT | RC | CD | CT |
| Time(min) | 0.04 | 0.2 | 2.8 | 0.82 | 4.02 | 16.25 | 5.29 | 36.5 | 87 |

## 3.2 Assessing Cluster Quality

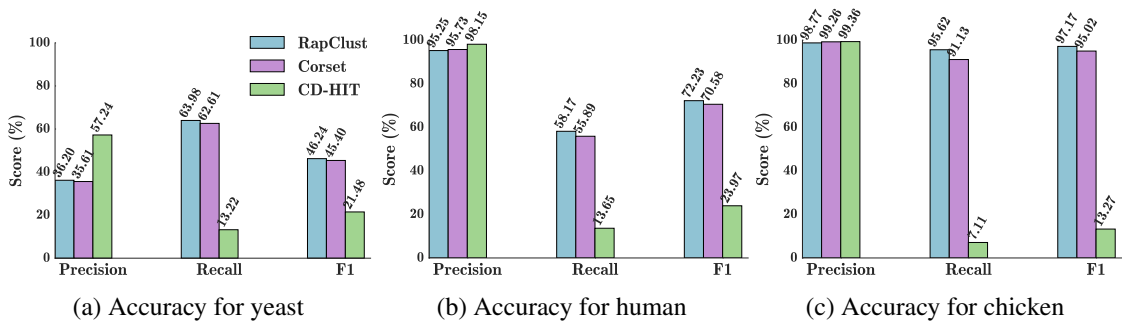

(a) Accuracy for yeast      (b) Accuracy for human      (c) Accuracy for chicken

Figure 2: The precision, recall, and F1-score of the `RapClust`, `Corset` and `CD-HIT` based clusterings, with respect to ground-truth annotations, on the yeast (a), human (b) and chicken (c) data sets.

We assessed the quality of the clusters obtained by the various tools using two different metrics. The ground truth labels for *de novo* assembled contigs were taken from (4), and the process of obtaining these labels is described in (5). It is important to note that not all contigs can be labeled with an annotated gene, and unlabeled contigs were omitted when computing the metrics below.

First, we considered the precision and recall metrics used by Davidson and Oshlack (5). Here, pairs of contigs are classified based on whether their cluster labels match their annotated gene labels. Specifically, a pair of contigs labeled with the same gene is considered a true-positive (TP) if the pair is co-clustered and a false-negative (FN) if the contigs are placed in separate clusters. Likewise, if a pair of contigs labeled with different genes is placed into the same cluster, it is considered a false-positive (FP) and if they are placed in different clusters, it is considered a true-negative (TN). From these pairwise counts, the precision and recall can be computed as Precision = TP/TP+FP and Recall = TP/TP+FN. A higher precision signifies that, when contigs are co-clustered, they are more likely to have originated from the same underlying gene. A higher recall, on the other hand, suggests that more contigs originating from the same gene tend to be co-clustered. Typically, precision and recall are competing objectives, as *over clustering* will improve recall but harm precision while *under clustering* will improve precision but harm recall. Commonly, the F1-Score = 2 (Precision·Recall/Precision+Recall) is used as a single metric to summarize performance in terms of both the precision and recall. Figure 2 shows the accuracy of all three tools on the three assemblies in terms of Precision, Recall and F1-Score. `Corset` and `RapClust` generate similar clusters, with `RapClust` generally yielding slightly higher recall than `Corset`. `CD-HIT`, however, tends to do a much poorer job at trading off between these competing objectives. It provides similar (or, in case of the yeast data, higher) precision to `Corset` and `RapClust`, but the resulting clusters exhibit much lower recall.

Table 4: The variation of information between contig to gene mapping using genome-based mapping approach and the clusters generated using `RapClust`, `Corset` and `CD-HIT`. For each assembly, the clustering producing the lowest variation of information with respect to the true clustering is set in bold; `RapClust` achieves the lowest VI on all assemblies.

|         | RapClust | Corset | CD-HIT |
|---------|----------|--------|--------|
| Chicken | **0.127** | 0.191 | 2.01 |
| Human   | **0.712** | 0.735 | 1.24 |
| Yeast   | **0.176** | 0.178 | 0.216 |

Second, we considered how similar the clusters obtained using the different methods are to the ground truth clustering, which groups together all contigs labeled with the same gene. To assess this similarity, we used the variation of information (VI) (13). The VI is defined over a pair of clusterings, and quantifies the information lost and gained when moving from one clustering to the other. It allows one to measure how similar two clusterings are, regardless of the specific names or labels chosen for the clusters. The lower the VI between a pair of clusters, the more similar they are. To compute the VI between the true clustering $C_T$ and that obtained by a particular method $C_M$, we discarded all contigs in $C_M$ that are not labeled with a gene name, while retaining the clustering relations among the remaining contigs. If any contigs that correspond to annotated genes do not exist in $C_M$ (since they may be discarded, e.g., by the read count filter described in section 2.3), we considered them to come from a single cluster, which is given a new label. We called the resulting clustering $C_{M'}$. Hence, $C_T$ and $C_{M'}$ are defined over the same set of contigs, and the similarity between them can be computed directly using the VI. The VI results are presented in Table 4. `RapClust` and `Corset` seem to yield similar results, with `RapClust` obtaining a slightly lower (better) VI. However, we observed a *marked* difference between these two and `CD-HIT`, whose clusters exhibit a substantially larger VI, especially on the human and chicken data.

### 3.3 Differential Gene Expression

We also tested the ability to recover gene-level differential expression using the clusterings produced by the different methods. *De novo* transcriptome assemblers have a tendency to produce many (often fractured) contigs. This tends to confound downstream differential expression analyses, due, in part, to the difficulty of quantifying fractured or incorrectly assembled contigs, and due, in part, to the potentially large number of extra statistical tests being performed (that must be corrected for). By estimating expression, and per-forming differential expression testing at the cluster level, one might expect to simultaneously reduce the multiple hypothesis testing burden and "average out" some of the mistakes made in contig-level abundance estimation.

For gene-level differential expression analysis, we compared the genes called as differentially expressed under each of the different clusterings versus the genes detected as differentially expressed under the true contig-to-gene mapping (again, note that not all contigs are annotated with a gene label). We generated "ground truth" gene-level abundance estimates based on the contig-level abundance estimates computed by `Sailfish`, and the true contig-to-gene mapping. Using the tximport (24) R package, we loaded the `Sailfish` expression estimates, aggregated them to the gene level, and prepared them for use with DESeq2 (12). A 2-condition differential expression test was performed in each data set; for human this was in between the conditions with and without the HOX1A knockdown, for yeast it was in the batch and chemostat growth conditions, and for chicken it was in the male and female samples (we collapsed the different tissues within each sex). We then obtained corrected p-values for the hypothesis that each gene is differentially expressed across the conditions we considered. We considered genes having a corrected p-value less than or equal to 0.05 as differentially expressed. We estimated differential expression under the `RapClust` and `CD-HIT` clusterings in the same manner, where the quantification estimates were held fixed, but the true contig-to-gene mapping was replaced with the contig-to-cluster mapping produced by these tools. For `Corset`, a count matrix is directly provided that was used as input to DESeq2.

As a metric of comparison, we examined the rate at which true positive differentially expressed genes were recovered versus the rate at which false positive differentially expressed genes were called. Each predicted cluster was labeled with the union of all of the genes with which its constituent contigs were labeled. We sorted the list of clusters by p-value, and processed them in the following manner: when we encountered a cluster, we intersected its set of labeled genes with the set of truly differentially expressed genes. Any genes that had not already been encountered in a more highly-ranked cluster were considered as true positive predictions. Likewise, any genes that appeared in the cluster, but which did not occur in the true set of differentially expressed genes were considered as false positives (if the genes had not already been encountered in a more highly-ranked cluster). We stopped processing the clusters once their adjusted p-value exceeded 0.05 (since, under common threshold, such clusters would likely not be considered to be differentially expressed). For the true and false positive predictions we encountered, the associated negative p-value of the associated cluster was used as the corresponding "score". The ROC curves were generated using the CROC (28) Python package.

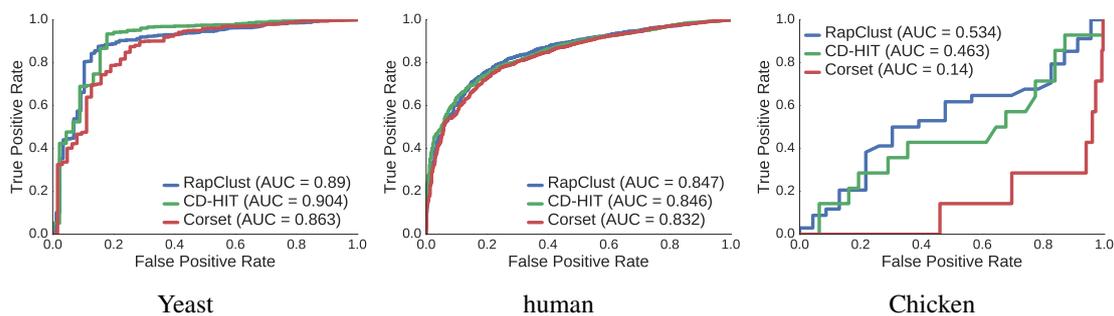

Yeast　　　　　　　　　　　　human　　　　　　　　　　　　Chicken

Figure 3: ROC curves showing the recovery rate of `RapClust`, `Corset`, and `CD-HIT`'s clusters in recovering differentially expressed genes in each data set.[1]

Overall, poor precision or recall in terms of clustering may lead to detection of fewer truly differentially expressed genes, spurious identification of differential expression, or weak statistical evidence for differential expression. Figure 3 shows that the rate at which `RapClust` recovers true positives versus false positives is higher than that of `CD-HIT` in 2 of the 3 assemblies, and is higher than that of `Corset` in all the assemblies, as represented by the respective area under the curves. The benefit of `RapClust` is particularly apparent in the chicken assembly. Since the quantification results produced by `Sailfish` tend to be reasonable, even when the clustering is fairly poor, this may explain the relatively good performance of the `CD-HIT` clustering in yeast (relative to the rather poor quality of the clustering, as evaluated in Section 3.2).

---

[1] The rather poor performance of `Corset`, under this particular test, the Chicken assembly seems *not* to be the result of a poor clustering, and is explored further in Table S1 and Figure S1.

## 4  Conclusion and Discussion

We have presented a fast and accurate methodology for the data-driven clustering of *de novo* transcriptome assemblies. By making explicit the connection of the method of `Corset` (5) and the fragment equivalence classes that have recently been proven useful in the development of accurate and fast RNA-seq quantification tools (3, 17, 18), we have demonstrated how state-of-the-art transcript clustering results can be obtained much more quickly than is possible with existing tools. Furthermore, when working directly in the compact and efficient representation of fragment equivalence classes, this clustering requires only a marginal computational and storage cost beyond what is already required for *quasi-mapping*-based transcript-level quantification. We have implemented this approach in the open-source software `RapClust`.

There are many interesting directions for future work on this problem. Here, we mention only the most obvious. First, we believe that the quality of the resulting clusters could be improved through a data-driven selection of the appropriate cutoff parameters. Currently, we have directly adopted the parameters suggested in (5), which are selected partly independent of the underlying data. Preliminary experiments have suggested that one might be able to obtain substantially better clusters by selecting the transcript count cutoff, an edge-weight cutoff, and the log fold-change likelihood cutoff in a manner that is more data-adaptive. However, automatically selecting these cutoffs in a general, yet rigorous fashion is a topic for future work. Another potential improvement on the current methodology would be to adopt a more robust log fold-change test, that may be more accurate in separating contigs that do not originate from the same gene. `Sailfish` is capable of producing not only transcript-level abundances, but also estimates of the variance of each predicted abundance via posterior Gibbs sampling or bootstraps. This variance information can be incorporated into the estimates of log fold-change differences to allow for increased precision in separating potential paralogs. While the existing method works well in the completely *de novo* context (i.e. even when the genomes or transcriptomes of closely related organisms may not be available), integrating homology information, when available, has the potential to improve the clustering results (and provide meaningful biological annotations for the clusters). The best way to integrate this information is an exciting direction for future work. Finally, we believe that sequence-level comparison and analysis of the resulting clusters of contigs could reveal important information about the nature of the transcripts present in the samples. For example, one could imagine "reverse-engineering" the splicing patterns present in the transcripts occurring in the same cluster. This would allow one to build a virtual gene model, even in the completely *de novo* context, which could then be used downstream, such as for differential splicing analyses.

# Supplemental Materials: Accurate, Fast and Lightweight Clustering of *de novo* Transcriptomes using Fragment Equivalence Classes

Table S1: The number of contigs retained, and corresponding number of clusters generated, by each of the methods tested for each data set. The number of contigs retained by `RapClust`, and the number of clusters it produces, generally reside between the corresponding values of `Corset` and `CD-HIT`.

| Chicken | CD-HIT | Corset | RapClust |
|---|---|---|---|
| # contigs | 331,450 | 181,333 | 301,913 |
| # clusters | 281,686 | 91,653 | 204,189 |
| Human | CD-HIT | Corset | RapClust |
| # contigs | 107,389 | 69,107 | 79,921 |
| # clusters | 90,115 | 43,663 | 52,575 |
| Yeast | CD-HIT | Corset | RapClust |
| # contigs | 7353 | 4145 | 4457 |
| # clusters | 7117 | 3796 | 4071 |

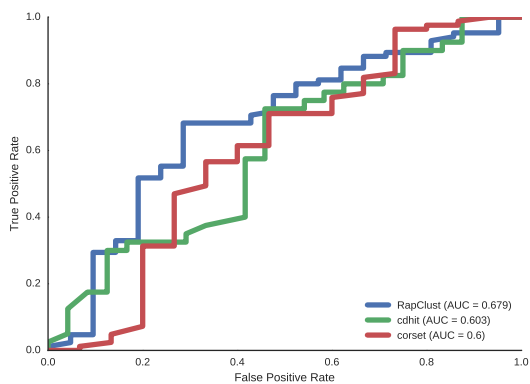

Figure S1: ROC curves showing the recovery rate of `RapClust`, `Corset`, and `CD-HIT`'s clusters in recovering differentially expressed genes in the Chicken dataset. Here, unlabeled contigs are removed from consideration in the predicted clusters, and the result labeled `Corset` uses the clustering obtained by `Corset`, but couples them with the abundance estimates produced by `Sailfish`. Under these testing conditions, all methods demonstrate a more favorable recovery curve (as expected, since unlabeled contigs are removed from the clusters as in the "ground truth" expression estimates), but the accuracy of `Corset`, in particular, shows the greatest improvement. This suggests that the clusters obtained by `Corset` are accurate, and that the performance observed in Figure 3 is likely the result of `Corset`'s substantially lower number of overall clusters on this dataset (see Table S1) and the differences between simple cluster-level read counting and contig-level abundance estimation.